\documentclass[conference]{IEEEtran}
\usepackage{color,soul}
\usepackage{multirow}
\usepackage{graphicx}
\usepackage{url}
\ifCLASSINFOpdf
\else
\fi
\begin{document}
%
\title{Remote Rendering for Virtual Reality: performance comparison of multimedia frameworks and protocols}

\author{
\IEEEauthorblockN{Daniel Mejías\IEEEauthorrefmark{3}, Inhar Yeregui\IEEEauthorrefmark{3}, Roberto Viola}
\IEEEauthorblockA{Fundación Vicomtech\\
Basque Research and Technology Alliance\\
San Sebastián, Spain\\
Email: \{damejias,iyeregui,rviola\}@vicomtech.org}
\IEEEauthorblockA{\IEEEauthorrefmark{3}PhD Candidate at UPV/EHU}
\and
\IEEEauthorblockN{Miguel Fernández\IEEEauthorrefmark{1}, Mario Montagud\IEEEauthorrefmark{1}\IEEEauthorrefmark{2}}
\IEEEauthorblockA{\IEEEauthorrefmark{1}i2CAT Foundation, Barcelona (Spain)\\
\IEEEauthorrefmark{2}Universitat de Val\`{e}ncia, Val\`{e}ncia (Spain)\\
Email: \{miguel.fernandez, mario.montagud\}@i2cat.net}}


%


\maketitle

\begin{abstract}
The increasing complexity of Extended Reality (XR) applications demands substantial processing power and high bandwidth communications, often unavailable on lightweight devices.
Remote rendering consists of offloading processing tasks to a remote node with a powerful GPU, delivering the rendered content to the end device. The delivery is usually performed through popular streaming protocols such as Web Real-Time Communications (WebRTC), offering a data channel for interactions, or Dynamic Adaptive Streaming over HTTP (DASH), better suitable for scalability.
Moreover, new streaming protocols based on QUIC are emerging as potential replacements for WebRTC and DASH and offer benefits like connection migration, stream multiplexing and multipath delivery.
This work describes the integration of the two most popular multimedia frameworks, GStreamer and FFmpeg, with a rendering engine acting as a Remote Renderer, and analyzes their performance when offering different protocols for delivering the rendered content to the end device over WIFI or 5G. This solution constitutes a beyond state-of-the-art testbed to conduct cutting-edge research in the XR field.
\end{abstract}

\begin{IEEEkeywords}
Extended Reality, Cloud/Edge Rendering, Multimedia streaming, Volumetric multimedia processing.
\end{IEEEkeywords}

%
\IEEEpeerreviewmaketitle

\section{Introduction}

In recent years, Extended Reality (XR) technology has grown in popularity as XR applications and use cases are increasing thanks to progress in network capability and computing hardware. This growth comes at the cost of increasing demand for processing powers, as XR applications execute complex processing tasks requiring Graphics Processing Units (GPUs) to perform them in real time.

Nevertheless, the processing power is not always available on the end devices. Virtual Reality (VR) headsets, smartphones and tablets, as well as lightweight laptops, are not provided with the necessary hardware acceleration to run XR applications. For such a reason, Remote Rendering solutions are being used as an alternative to the local processing on the end device. Heavy volumetric video rendering tasks are offloaded to a remote GPU-equipped computing node, which delivers only the processing result to the end device \cite{xiao2020cloud,yeregui2024edge}.
The result can be a 2D directive or a 360$^{\circ}$ video stream, associated with an audio stream.

Current solutions for Remote Rendering deliver the content to the end device using Web Real-Time Communications (WebRTC) \cite{webrtc}, while exploiting its data channel to exchange information concerning user interactions with the VR scene \cite{casasnovasexperimental,feng2024application}. WebRTC protocol is a perfect fit for its real time latency that allows the interaction information to be delivered as fast as possible. Nevertheless, it has an intrinsic limitation when high scalability is requested. Employing a Hypertext Transfer Protocol (HTTP) Adaptive Streaming (HAS) protocols, e.g., such as Dynamic Adaptive Streaming over HTTP (DASH) \cite{iso23009}, enables the use of Content Delivery Networks (CDNs) to increase the number of connected users. HAS protocols constitute the best solutions when the scalability is more important than the interaction capabilities, as the latter are limited only to rotation within the VR scene and only when 360$^{\circ}$ video is employed \cite{podborski2017virtual}.

Furthermore, the rise of 
QUIC \cite{rfc9000} as the transport protocol for HTTP/3 \cite{trevisan2021measuring} has the potential to be a game changer for future multimedia communications. Its native features such as session migration, stream multiplexing and multipath delivery make it suitable for user mobility or flexible network reconfiguration scenarios. Consequently, new streaming protocols, such as Real-time Transport Protocol (RTP) over QUIC (RoQ) \cite{ietf-avtcore-rtp-over-quic-11} and Media over QUIC (MoQ) \cite{ietf-moq-transport-06}, are proposed and some initial implementations are already available and tested \cite{engelbart2021congestion,gurel2023media}. If these protocols reach a mature development and become widely used, they could replace WebRTC and DASH, as they could ideally combine real time capabilities with scalability supported by CDN infrastructure.

The contributions of this work are as follows:
\begin{itemize}
    \item A full-fledged Remote Rendering pipeline capable of offloading from headsets the processing tasks required for rendering VR experiences, generating a 360$^{\circ}$ video stream compatible with any lightweight client device, while maintaining responsive interactions.
    \item A performance comparison among two prominent open-source media streaming frameworks, GStreamer \cite{gst} and FFmpeg \cite{ffmpeg}, including their advantages and disadvantages when integrated into XR services.
    \item An analysis of the latency achieved by standard multimedia protocols—WebRTC, DASH, Low Latency DASH (LL-DASH) \cite{lldash}, QUIC and MoQ—when delivering remotely rendered content.
\end{itemize}

Even if Remote Rendering is relevant and applicable to any XR scenario, this work focuses primarily on VR to maximize clarity and depth. However, the contributions made can effectively improve a variety of immersive experiences in the broader XR landscape.

 
\section{Background and Related work}

Different authors have addressed the need for remote rendering in various applications, such as scientific simulation, XR, training, computer-aided design (CAD), and cloud gaming \cite{4069234}. Studies have shown that executing rendering tasks remotely enables users to access complex virtual scene environments from resource-constrained devices \cite{10316520}. However, this approach introduces challenges related to real-time data transmission. To mitigate these issues, researchers have explored the use of communication protocols optimized for multimedia data. For instance, the RTP protocol has been widely adopted in image capture and videoconferencing applications due to its ability to handle real-time streaming \cite{10271910,10314778}. Nevertheless, significant challenges remain, particularly in terms of adaptability to different types of content, scalability across various devices, and efficiency in low-bandwidth networks, underscoring the need for further research in this area \cite{shi2010high,shi2012real}.

Modern frameworks for virtual scene development engines support remote communication using RTP-based protocols.
WebRTC is one of the most versatile solutions for real-time communication and is integrated into some development engines.
Unity and Unreal Engine are widely used engines for video games and interactive 3D applications, each with distinct characteristics, but both are highly versatile due to their integration with multiple platforms \cite{10808154,10271910,10608249}. Unity offers Universal Render Streaming\footnote{https://docs.unity3d.com/Packages/com.unity.renderstreaming}, which enables remote rendering with various graphical features, including encoding, dynamic bitrate adjustment, and communication via  WebRTC. Similarly, Unreal Engine provides Pixel Streaming\footnote{https://www.streampixel.io/}, which allows rendering and transmitting video output to the client through WebRTC communication.


However, in some cases, WebRTC may not be the most optimal choice, as firewall restrictions depending on network conditions may require specific port openings or the use of STUN/TURN servers, introducing additional latency. Additionally, the QUIC protocol may offer better performance in terms of reduced negotiation time and lower latency \cite{10808154}. Currently, there is no QUIC-based development in rendering engines, although it is an emerging and rapidly growing protocol. Another important requirement is the ability to achieve scalability and video adaptability according to network conditions using adaptive bitrate protocols such as DASH.
Therefore, the need for systems capable of supporting multiple communication protocols becomes even more critical, as well as the development of protocols adapted to the continuous evolution of networks and applications. Integrating a virtual scene development engine with frameworks such as GStreamer\cite{gst} or FFmpeg\cite{ffmpeg} can enable the use of different protocols, allowing greater adaptability to user requirements.

At the user level, the renderer must be accessible from any end device while maintaining the same quality of experience and low-latency characteristics. For this reason, the most optimal implementation is the use of a web browser as the rendering client. Many real-time communication protocols support web-based player development, enabling seamless integration with remote rendering. Additionally, technologies such as WebXR\footnote{https://immersiveweb.dev/} facilitate 3D visualization on headsets and mobile devices.



\section{Multi-framework and Multi-protocol Remote Rendering solution}

\begin{figure*}[!t]
\centering
\includegraphics[width=0.80\textwidth]{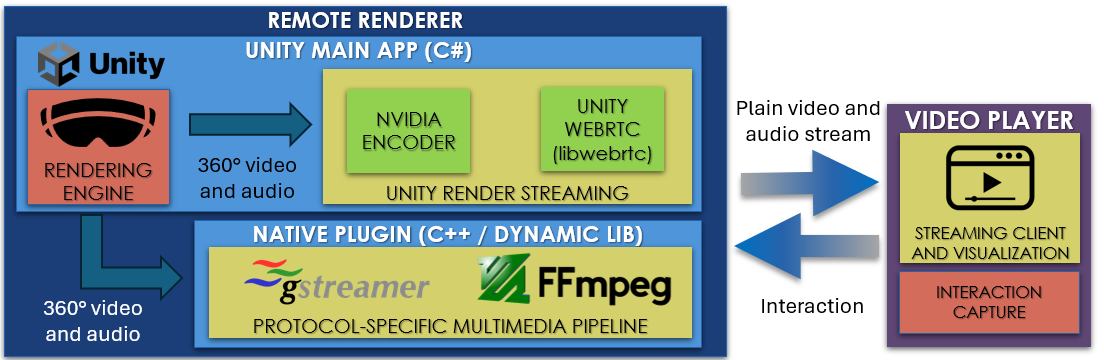}
\caption{General solution for multi-protocol Remote Renderer.}
\label{fig:general}
\end{figure*}

Figure \ref{fig:general} presents the proposed and implemented architecture of the Remote Renderer for VR experiences that communicates with the video player on the end device using different standardized streaming protocols.
The Remote Renderer consists of two main software components: a Unity main application, developed in C\#, and a native rendering plugin, developed in C++ and dynamically linked to the main application.

The Unity main application acts as the central hub, coordinating the interactions between the different components. It is based on the Unity rendering engine and the Unity Render Streaming (URS) plugin \cite{URS}. The former uses NVIDIA GPU capabilities to process heterogeneous content integrated in a VR scene, including audio and volumetric video, and generate rendered raw audio and 360$^{\circ}$ video streams.
The latter receives the raw streams, encodes them and generates a WebRTC stream to be delivered to the video player. NVIDIA encoder and \textit{libwebrtc} library \cite{libwebrtc} are employed for GPU encoding and WebRTC communication, respectively. The audio is encoded with OPUS codec, while the video with H.264 codec.
    
The native rendering plugin \cite{nativeplugin} consists of a custom Unity plugin to bridge the Unity main application and the multimedia framework in charge of creating a protocol-specific multimedia pipeline. Since no framework supports all the streaming protocols, two plugin versions are developed to integrate GStreamer and FFmpeg. The differences in implementation and supported protocols are described in the specific subsections.

Overall, the combination of URS, GStreamer and FFmpeg enables to generate WebRTC, DASH, LL-DASH, QUIC and MoQ streams. These standardized protocols have the advantage of enabling the reception, decoding and visualization by compliant video players without modifications. Adding user interaction instead requires introducing mechanisms for capturing the interaction on the end device, which are currently available only when using the URS to generate WebRTC.

In conjunction, this constitutes a beyond state-of-the-art testbed to conduct cutting-edge research in this field. It allows to find out pros and cons of different alternatives for media processing, streaming and interaction.

\subsection{GStreamer}

\begin{figure}[!t]
\centering
\includegraphics[width=0.48\textwidth]{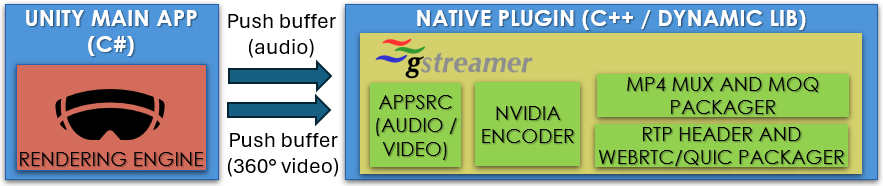}
\caption{Integration of GStreamer with the Remote Renderer.}
\label{fig:gstreamer}
\end{figure}

Figure \ref{fig:gstreamer} represents the integration of the GStreamer native plugin with the Unity main application and the operations performed by its internal multimedia pipelines. GStreamer is integrated by calling its native Application Programming Interfaces (APIs) to build and run the multimedia pipeline. Two GStreamer \textit{appsrc} \cite{gstappsrc} elements at the beginning of the pipeline allow pushing separately the 360$^{\circ}$ video and audio buffers into the pipeline. Then, the buffers are encoded with the support of NVIDIA encoders. The video is always encoded with H.264 codec, while the audio could use either OPUS or AAC. Depending on the protocol, the last operation within the pipeline is different. Current supported protocols and operations are as follows:
\begin{itemize}
    \item WebRTC: the encoded flows are packaged into a WebRTC stream. The resulting pipeline is an alternative to the URS WebRTC implementation and enables a direct comparison with it. 
    \item DASH: the encoded flows are muxed and split into MP4 segments, while a DASH Media Presentation Description (MPD) is generated. This pipeline enables higher scalability by design, even if it comes at a cost of higher latency compared to the WebRTC pipeline.
    \item QUIC: the encoded video is packaged into RTP streams \cite{rfc3550}. Differently from all the other protocols which belong to the application layer, this is simply a transport layer protocol. Since no application layer protocols based on QUIC are officially supported yet by GStreamer, this can be considered as a reference for future application layer implementations. On top of QUIC, a legacy RTP stream is employed. This must not be confused with RoQ which requires further adaptations of the RTP stream.
    \item MoQ: a fragmented MP4 (fMP4) muxer is employed for packaging the encoded stream. Then, an unofficial GStreamer element implements MoQ to deliver the fMP4 flow \cite{curley2024media}. Since this element is not included in stable and maintained GStreamer releases, it is less mature than other alternatives and does not support audio yet. However, it constitutes the only QUIC-based application layer protocol.
\end{itemize}

\subsection{FFmpeg}

\begin{figure}[!t]
\centering
\includegraphics[width=0.48\textwidth]{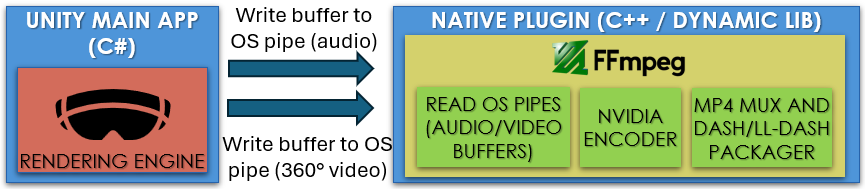}
\caption{Integration of FFmpeg with the Remote Renderer.}
\label{fig:ffmpeg}
\end{figure}

Figure \ref{fig:ffmpeg} represents the integration of the FFmpeg native plugin with the Unity main application and the operations performed by its internal multimedia pipeline. Differently from GStreamer, FFmpeg is integrated by calling its Command Line Interface (CLI) and using two Operating Systems (OS) pipes \cite{pipe} to separately write the 360$^{\circ}$ video and audio buffers from Unity and read them from the FFmpeg pipeline. Then, the buffers are encoded with the support of NVIDIA encoders. The video is encoded with H.264 codec, while the audio with AAC codec. For the last processing step, two alternatives are available depending on the protocol:
\begin{itemize}
    \item DASH: the encoded flows are muxed and split into MP4 segments, while a DASH MPD is generated. This is similar and directly comparable with GStreamer DASH pipeline. 
    \item LL-DASH: similar to the DASH pipeline, the segments and MPD are generated, but a fMP4 muxed is instead employed. It allows the reduction of MP4 fragment duration within the segments to lower the latency.
\end{itemize}

\section{Experimental results}

\begin{figure}[!t]
\centering
\includegraphics[width=0.48\textwidth]{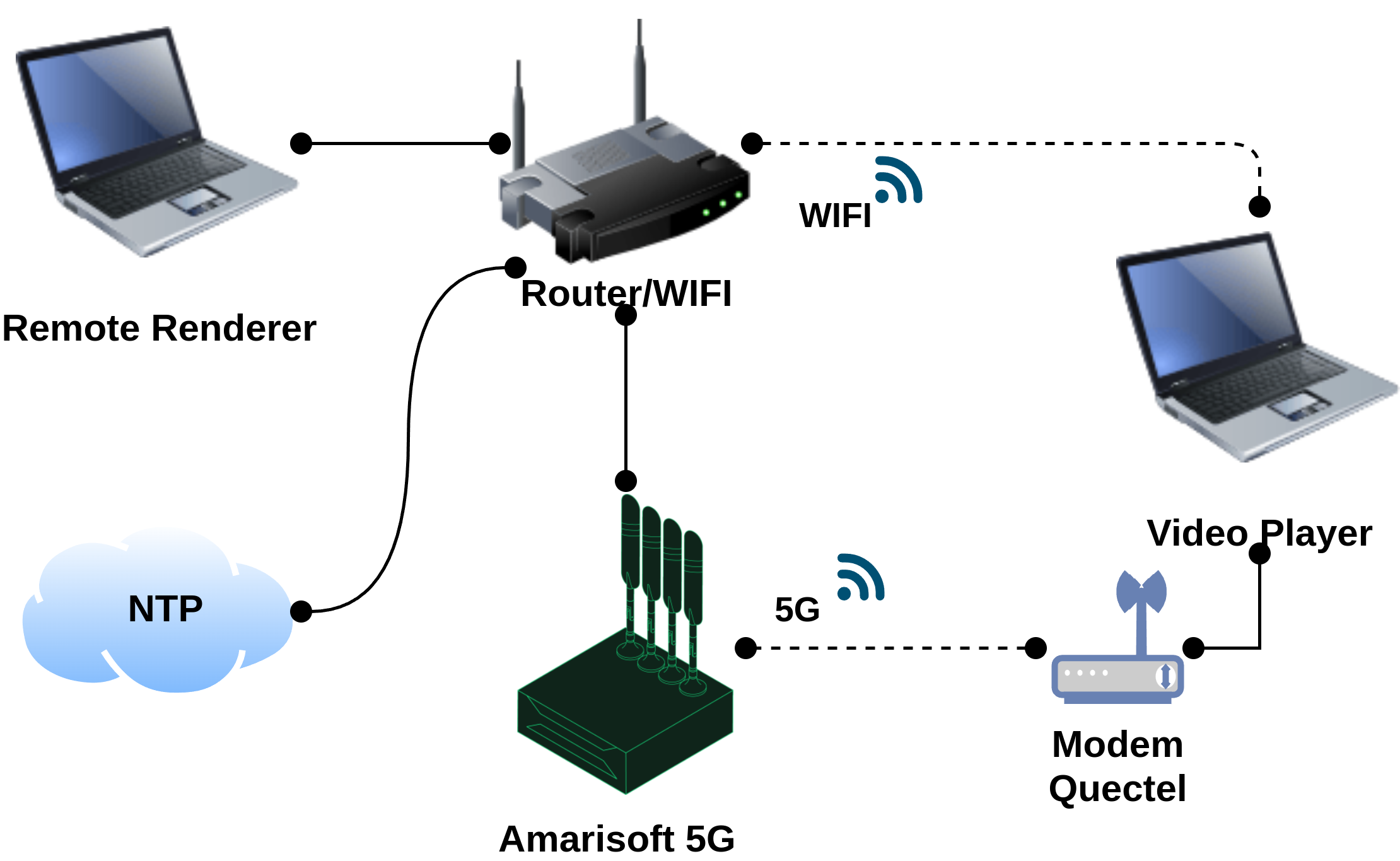}
\caption{Testbed for WIFI and 5G latency measurement.}
\label{fig:testbed}
\end{figure}

The testbed in Figure \ref{fig:testbed} is employed to compare the multimedia frameworks and protocols integrated into the Remote Rendering solution. Its specifications are as follows:
 
\begin{itemize}
    \item Remote Renderer: a laptop acting as the rendering node with an Intel i7-13650HX Central Processor Unit (CPU), 16GB Double Data Rate 5 (DDR5) Random-Access Memory (RAM)
    and NVIDIA RTX4060 GPU. 
    \item Router WIFI: a WIFI 6 compatible router employing the 5GHz band for communication between the Remote Renderer and the video player. Alternatively, this equipment can route the traffic to the 5G network instead of WIFI.
    \item AMARI Callbox Mini \cite{AMARI}: equipment with 5G Core and Radio Access Network (RAN) implementing the 3rd Generation Partnership Project (3GPP) Release 15 \cite{rel15}.
    It is configured to use the N78 band, 20 MHz bandwidth, 30 MHz subcarrier spacing, max modulation of 256 Quadrature Amplitude Modulation (QAM), 2 antennas for downlink and 1 for uplink.
    \item Quectel 5G modem: Quectel RM500Q modem implementing 3GPP Release 15 and connected to the video player equipment via USB.
    \item Video Player (laptop): the video player is displayed from a web interface or an application (only for QUIC). Any computer with basic capabilities can be used. This laptop has an internal WIFI 6 adapter and the driver installed for the Quectel modem.
    \item Network Time Protocol (NTP): the router provides an Internet connection that allows the Remote Renderer and the Video Player to synchronize to a public NTP server for latency testing.
\end{itemize}


\begin{figure}[!t]
\centering
\includegraphics[width=0.5\textwidth,clip,keepaspectratio]{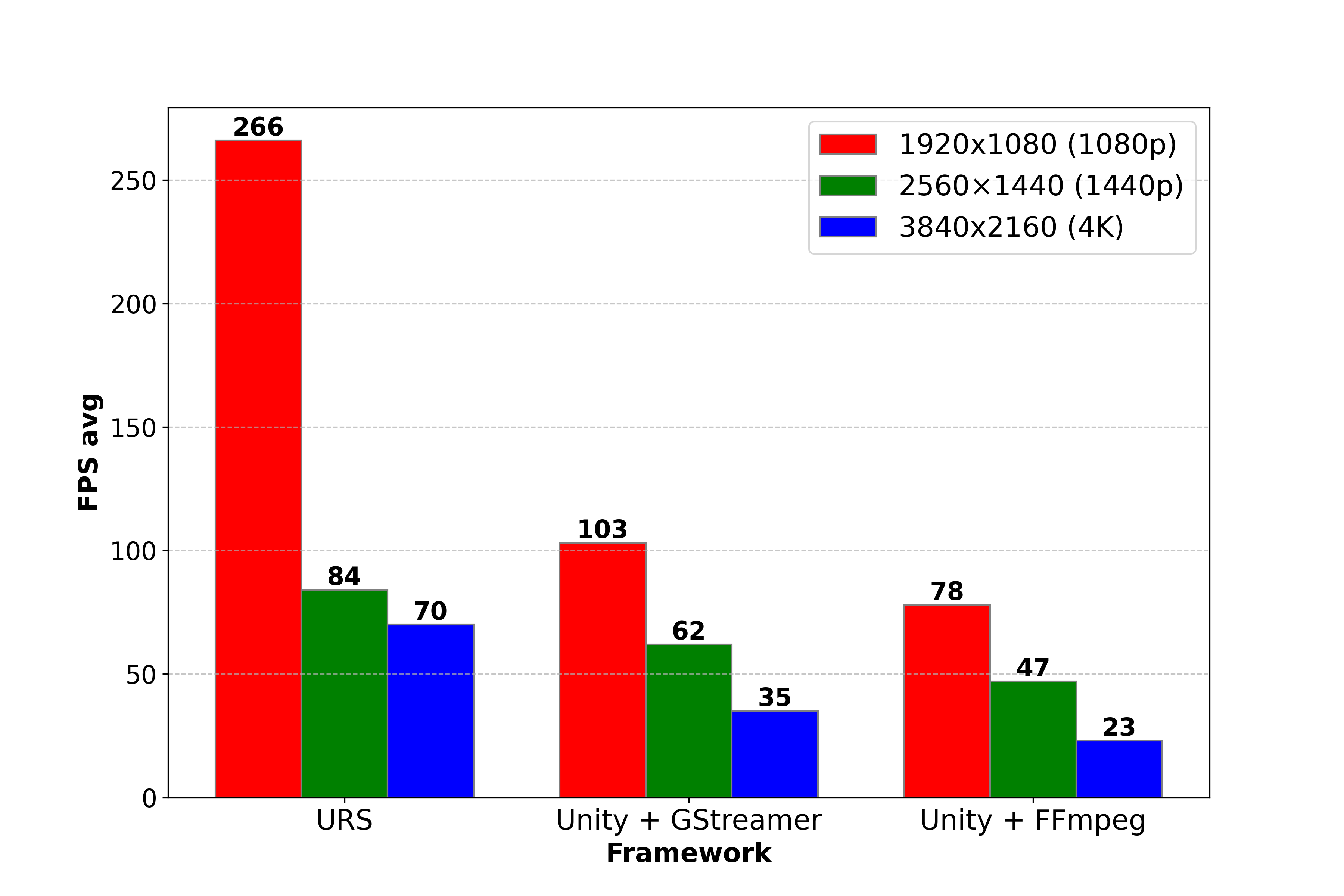}
    \caption{Maximum fps for each framework.}
    \label{fig:fps}
\end{figure} 

\begin{table}[!t]
\caption{Latency over WIFI with video stream at 1080p/60fps/10Mbps.}
\centering
\bgroup
\def\arraystretch{1.2}
\setlength\tabcolsep{2.5pt} 
\label{tab:latency}
\begin{tabular}{|c|c|c|c|c|}
\hline
\multirow{2}{*}{\textbf{Protocol}} & \textbf{Sender} & \textbf{Receiver} & \textbf{Latency$_{avg}$} & \textbf{Latency$_{dev}$} \\ 
& \textbf{framework} & \textbf{framework} & \textbf{(ms)} & \textbf{(ms)} \\
\hline
\multirow{4}{*}{WebRTC} & \multirow{2}{*}{URS} & URS player & \multirow{2}{*}{82} & \multirow{2}{*}{12} \\
& & \cite{URS} & & \\
\cline{2-5}
& Unity + & gstwebrtc-api & \multirow{2}{*}{219} & \multirow{2}{*}{22} \\
& GStreamer & \cite{gstwebrtcapi} & & \\
\hline
\multirow{3}{*}{DASH} & Unity + & Dash.js \cite{dashjs} & 9399& 13\\
& GStreamer & & & \\
\cline{2-5}
& Unity + FFmpeg & Dash.js \cite{dashjs} & 6183& 11\\
\hline
LL-DASH & Unity + FFmpeg & Dash.js \cite{dashjs} & 4873& 13\\
\hline
QUIC & Unity + & GStreamer & \multirow{2}{*}{248} & \multirow{2}{*}{2} \\
(RTP) & GStreamer & pipeline & & \\
\hline
\multirow{2}{*}{MoQ} & Unity + & MoQ player & \multirow{2}{*}{159} & \multirow{2}{*}{12} \\
& GStreamer & \cite{curley2024media} & & \\
\hline
\end{tabular}
\egroup
\end{table}

\begin{table}[!t]
\caption{Latency over 5G with video stream at 1080p/60fps/10Mbps.}
\centering
\bgroup
\def\arraystretch{1.2}
\setlength\tabcolsep{2.5pt} 
\label{tab:latency2}
\begin{tabular}{|c|c|c|c|c|}
\hline
\multirow{2}{*}{\textbf{Protocol}} & \textbf{Sender} & \textbf{Receiver} & \textbf{Latency$_{avg}$} & \textbf{Latency$_{dev}$} \\ 
& \textbf{framework} & \textbf{framework} & \textbf{(ms)} & \textbf{(ms)} \\
\hline
\multirow{4}{*}{WebRTC} & \multirow{2}{*}{URS} & URS player & \multirow{2}{*}{129} & \multirow{2}{*}{11} \\
& & \cite{URS} & & \\
\cline{2-5}
& Unity + & gstwebrtc-api & \multirow{2}{*}{253} & \multirow{2}{*}{16} \\
& GStreamer & \cite{gstwebrtcapi} & & \\
\hline
\multirow{3}{*}{DASH} & Unity + & Dash.js \cite{dashjs} & 10227& 14\\
& GStreamer & & & \\
\cline{2-5}
& Unity + FFmpeg & Dash.js \cite{dashjs} & 6519& 11\\
\hline
LL-DASH & Unity + FFmpeg & Dash.js \cite{dashjs} & 5084& 19\\
\hline
QUIC & Unity + & GStreamer & \multirow{2}{*}{293} & \multirow{2}{*}{31} \\
(RTP) & GStreamer & pipeline & & \\
\hline
\multirow{2}{*}{MoQ} & Unity + & MoQ player & \multirow{2}{*}{194} & \multirow{2}{*}{20} \\
& GStreamer & \cite{curley2024media} & & \\
\hline
\end{tabular}
\egroup
\end{table}

The development involves the rendering of a Unity-generated virtual scene into a plain video representation, to be later encoded and transmitted using different protocols. The scene consists of a test environment with a simple clock object showing the timestamp. This clock is necessary to assess the exact time when each frame is generated. We execute two performance tests: the first consists of measuring the maximum framerate achieved in rendering task, and the second is an end-to-end video latency test.


In the first test, the results are presented in Figure \ref{fig:fps}, which shows the average maximum framerate (frames per second or fps) processed by each framework, independently of the considered protocol. These values are calculated by counting the frames after the GPU encoding and writing them directly into a file. URS is outperforming GStreamer and FFmpeg as it has the best integration with Unity, being part of its official plugins. When using GStreamer and FFMPEG, the raw frames are extracted from the GPU to be injected into their pipelines. This operation is inefficient as later both frameworks inject the frames back into the GPU for encoding. In any case, the usage of GStreamer native APIs provides better performance than FFmpeg CLI, as passing through OS pipes is not necessary. FFmpeg CLI is still a valid solution for fast prototyping due to its easier integration, even if the framerate at 4k might not guarantee a smooth visual experience.

In the second test, the results are presented in Table \ref{tab:latency} and \ref{tab:latency2} present the latency achieved by each streaming protocol when delivering a 1080p/60fps video encoded at 10Mbps over WIFI and 5G, respectively. Since a high framerate allows to reduce the latency \cite{geris2024balancing}, 1080p is chosen for these tests as it is the only resolution that each framework can provide with at least 60fps. Except for QUIC which employs a GStreamer pipeline as a player (a QUIC stream is not playable within a browser without an application layer protocol), all the other protocols employ reference web players.
For encoding, H.264 was used with an Nvidia GPU, configured with a Group of Pictures (GOP) of 5 frames, a encoding bitrate of 10 Mbps, and a frame rate of 60 FPS in URS, GStreamer and FFmpeg.

All the protocols show better results over WIFI than 5G, ranging from -36\% with URS WebRTC to -4\% with LL-DASH. WebRTC has the lowest latency when implemented with URS, while the GStreamer WebRTC alternative increases significantly by +167\% over WIFI and +96\% over 5G. Again, this can be explained by the inefficient process of extracting the frames from the GPU. The results of the QUIC transport protocol are quite promising for future multimedia communication. QUIC (RTP) shows a higher latency but close to GStreamer WebRTC, being the difference 29ms over WIFI and 40ms over 5G. MoQ implementation is instead already outperforming GStreamer WebRTC, giving the best results after URS WebRTC.

As expected, DASH and LL-DASH implemented with FFmpeg present the highest latency. Both of them are configured to generate segments of 2 seconds, but fMP4 allows LL-DASH to have 0.5 seconds fragments. The LL-DASH latency is 21\% lower than DASH over WIFI and -22\% over 5G. 
The generation of DASH content with GStreamer does not fully comply with the \textit{DASH-IF} recommendations.
It is worth noticing that the current GStreamer DASH implementation does not generate a stream complainant with DASH Industry Forum recommendations.  As a result, we had to modify it to make it compatible with \textit{Dash.js} and submitted a merge request to GStreamer\footnote{\url{https://gitlab.freedesktop.org/gstreamer/gstreamer/-/merge_requests/7886}}.
However, it exhibits a latency increase of +156\% compared to FFmpeg DASH, as the GStreamer implementation might not have the same maturity as FFmpeg. 

Finally, Table \ref{tab:frameworks} provides an overview of key features to consider when selecting any of the studied frameworks for producing and delivering rendered experiences as standard multimedia streams. Each framework has its advantages and disadvantages, and the choice depends on the specific requirements of the developers, such as support for new protocols, live adaptation of streaming parameters, or hardware (HW) control. It is useful to consider these parameters to make the best decision on which framework to choose.

\begin{table}[!t]
\caption{Characteristics of studied multimedia frameworks.}
\centering
\bgroup
\def\arraystretch{1.2}
\setlength\tabcolsep{2.0pt} 
\label{tab:frameworks}
\begin{tabular}{|l|c|c|c|c|}
\hline
\textbf{Framework} & \textbf{Prototype} & \textbf{Protocols} & \textbf{Tune} & \textbf{HW control}  \\ 
\hline
\multirow{2}{*}{URS} & built-in Unity & single & few parameters & no \\
 & plugin & protocol & update on the fly & control \\
\hline
\multirow{2}{*}{GStreamer} & modular & lots of & most parameters & GPU/CPU \\
 & integration & protocols &  update on the fly & and threads \\
\hline
\multirow{2}{*}{FFmpeg} & command-line & lots of & some parameters & enable \\
 & integration & protocols & update on the fly & presets \\
\hline
\end{tabular}
\egroup
\end{table}

\section{Conclusion}

This paper has described an architecture capable of offloading the computational load of the VR headset by remotely rendering experiences with capable systems and streaming the video with different protocols to lightweight client devices.
In addition, a performance study of two leading open-source multimedia streaming frameworks, GStreamer and FFmpeg, has highlighted their respective strengths and weaknesses in XR service integration, comparing them to the built-in URS Unity plugin. Finally, the analysis of multimedia protocols, such as WebRTC, DASH, LL-DASH, QUIC and MoQ, has provided valuable insights into their latency performance in delivering remotely rendered content, facilitating efficient and seamless XR experiences.

The final contribution of this paper is to provide a comprehensive testbed implementation of a Remote Rendering solution to conduct cutting-edge research in this field, along with insightful results that reveal its usefulness.


\section*{Acknowledgment}
This research was supported by the SNS-JU Horizon Europe Research and Innovation programme, under Grant Agreement 101096838 for 6G-XR project, the Spanish MINECO and the European Union – NextGeneration EU, in the framework of the PRTR Call UNICO I+D 5G 2021, under Grant TSI-063000-2021-4 for 6G-Openverso-Holo project.
The work of Mario Montagud has been funded by MCIN/AEI/10.13039/501100011033 under Grant RYC2020-030679-I and by the ESF.



\bibliographystyle{IEEEtran}
\bibliography{IEEEabrv,main.bib}
%



\end{document}